# A Pixel Detector for Level-1 Triggering at SLHC

J. Jones, G. Hall, C. Foudas, A. Rose

Imperial College, London, UK.
John.Jones@imperial.ac.uk

*Abstract*

We report on preliminary design studies of a pixel detector for CMS at the Super-LHC. The goal of these studies was to investigate the possibility of designing an inner tracker pixel detector whose data could be used for selecting events at the First Level Trigger. The detector considered consists of two layers of $20\times50\times10\mu m^3$ pixels at very close radial proximity from each other so that coincidences of hits between the two layers amount to a track transverse momentum ($p_T$) cut. This cut reduces the large amount of low-momentum data expected at SLHC while keeping the tracking efficiency very high for the high-$p_T$ tracks. Preliminary results on the performance of such a detector are presented.

## I. OVERVIEW

The current design of CMS is based on the nominal beam luminosity $10^{34}cm^{-2}s^{-1}$. It is anticipated that after running for several years, both LHC and the detectors will be upgraded to operate at a luminosity of $10^{35}cm^{-2}s^{-1}$ [1]. This presents a great challenge both in terms of radiation hardness and the increased data rates that will have to be sustained by the detectors and their corresponding DAQ systems.

The increase in luminosity at SLHC presents two problems for the current CMS DAQ readout. Firstly, the increased particle density in the detector (which scales with the luminosity of the machine) will result in an approximately ten-fold increase in bandwidth requirements for the readout of data associated with a single bunch crossing. The second problem relates to the performance of the Level-1 (L1) Trigger in CMS. The current system searches events with isolated leptons/photons, large missing/transverse energy and jets, as well as muons from the outer detector. Tracker information does not currently contribute at this level. The increased particle density in SLHC degrades the performance of the L1 trigger algorithms significantly due to the lack of isolated trigger objects and the negligible gains achieved by increasing $p_T$ thresholds for the muon systems. Figure 1 shows the limited ability to further reduce the muon trigger rate as the $p_T$ threshold is increased. Only the inclusion of data from the tracker in the Higher Level Trigger is able to reduce this rate further.

The former problem can be dealt with by increasing the DAQ bandwidth by a factor of ten. This is not considered a serious problem because of the continuing developments in semiconductor technology. However the second problem can only be dealt with by including information from the tracker in the L1 trigger system; an increase in L1 trigger rate is not considered an acceptable solution.

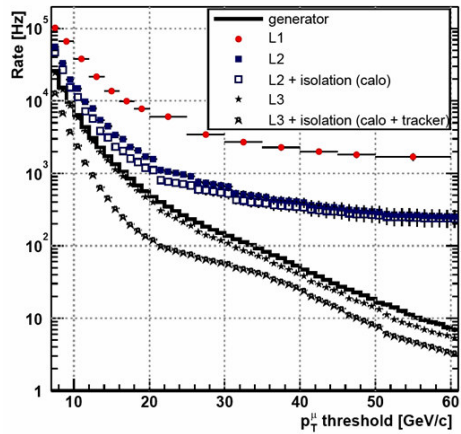

Figure 1: L1 single muon trigger rates for CMS [2]

The current CMS detector has two main parts. Broadly-speaking, the outer part of the tracker consists of many layers of microstrips of varying pitch, each connected to an APV25 readout chip [3]. This system is then linked to the outside DAQ system using analogue optical links. This analogue system is completely unsuitable for a contribution to L1 triggering, as zero-suppression for this system occurs off-detector on the Tracker Front End Driver, and therefore the time required for readout exceeds the Level-1 Trigger latency. The inner part of the tracker consists of three layers of pixels of pitch $100\times150\times300\mu m^3$. Unlike the APV25, the pixel ReadOut Chip (ROC) does perform zero-suppression [4], but it cannot contribute fully to L1 triggering in its present form as even the zero-suppressed data readout time is still too great to satisfy the Level-1 latency requirement.

### A. Tracker Contribution to Level-1 Triggering

Apart from jet vetoing by multiplicity, the simplest useful tracking contribution is a stub from two consecutive barrel layers. The stub can be used in coincidence with the outer detector to indicate whether the hit in the outer detector was caused by a high-$p_T$ particle. The quality of the stub (i.e. whether the hits are matched correctly between the two layers) is dependent on the layer separation; for layer separations of greater than a centimetre (see Figure 2), tracks from different events will overlap, producing a large number of track combinatorials during reconstruction. Therefore a 'standard'-spaced pixel detector would require 3-4 layers to provide a useful contribution. Implementing this in a detector upgrade is considered an impractical and expensive approach (both in terms of financial cost, power requirements, the requirement of inter-layer data transfer for the new system and the final rate of data flow out of the detector).

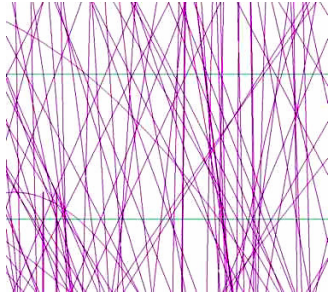

Figure 2: Track overlap in y-z plane (detector coordinates – see Fig. 3). 1cm layer separation is denoted by the two cyan lines. Note the significant overlap of tracks between these two layers, which will hinder tracker reconstruction.

### B. Stacked Tracking

An alternative approach to the combinatorial problem involves bringing two pixel layers together so that they are separated by approximately 1-2mm. The combinatorials then become manageable; even the limited knowledge of the interaction point is sufficient to make a 1:1 match between many of the hits in the two layers. This enables fast reconstruction using simple binning techniques, which could be implemented in an FPGA off-detector or a radiation-hard ASIC on-detector.

The basic layout of a stacked pixel detector is shown in Figure 3. If 100% signal efficiency is required, the arrangement can be made hermetic by overlapping the stacks in a similar way to that used in the current tracker.

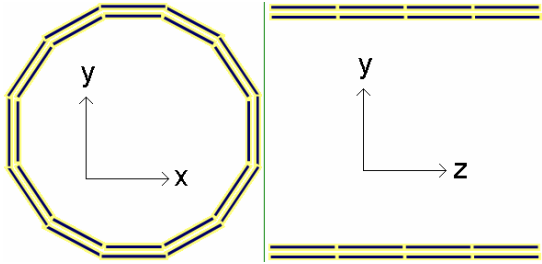

Figure 3: Basic layout of a flat stacked tracker (not to scale). Left is an x-y view, right is a y-z view.

### C. The Tracker Data-Rate Problem

The expected data rate for a binary pixel system at Super-LHC can be extrapolated from the occupancy of the pixel system at LHC. A rough calculation yields a value of approximately 4 hits per $(1.28\text{cm})^2$ at a layer radius of 10cm (full simulation yields a consistent but slightly lower number [5]). If one assumes a 16-bit pixel coding scheme, a naïve value for the data rate can be calculated as $3.125\text{Gb/cm}^2/\text{s}$. One must also include a coding scheme for the optical links (e.g. 8b10b, Hamming code) and a margin for additional coding information in the data stream. A very rough final number would then be **5Gb/cm$^2$/s**. This may be an overestimate, but it is still well beyond currently available link technology in radiation-hard form, and would result in large cabling and power requirements for the new detector.

### D. A Geometrical $p_T$ Cut

If one wishes to reduce the data rate from the new detector below that produced by a zero-suppressed binary readout, a novel method is required to filter the data. This new technique must necessarily discard real hit data. Collisions at SLHC produce a huge number of low-$p_T$ (<0.8GeV) particles that occupy the pixel detector but do not even reach the calorimeter because of the bending power of the 4T magnetic field (see Figure 4). The ideal solution for data rate reduction would be to filter these tracks from the data set.

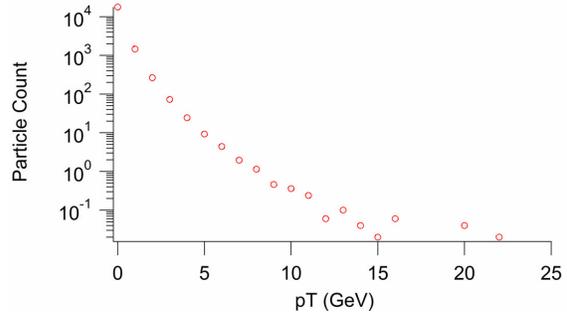

Figure 4: $p_T$ distribution of minimum bias charged particles in CMS (bunch-crossing-averaged); produced by superposing 100 minimum bias events / crossing generated by Pythia 6.2772 via CMKIN 4.2. The discontinuities at greater $p_T$ are a result of limited statistics.

The traditional approach to $p_T$ measurement of a charged particle track involves measuring the sagitta of the track as it travels through several layers of tracking detector. The process of reconstruction in this case involves large-scale communication between different detector layers, and uses relatively slow multiple-pass reconstruction methods to eliminate track combinations (i.e. Kalman filtering).

An alternative approach involves measuring the track crossing angle orthogonal to a layer's surface. This is directly related to the transverse momentum of the charge particle; the highest-$p_T$ tracks will cross almost orthogonal to the surface, whereas low-$p_T$ tracks will cross at a wider angle. The interesting feature of this method for a stacked tracker is that the rφ distance travelled between two sensors in a stack is of a similar size to the pitch of a single pixel. Hence by performing a nearest-neighbour search in the inner sensor of a stack using a seed hit in the outer sensor, one can isolate particles with a high transverse momentum (see Figure 5).

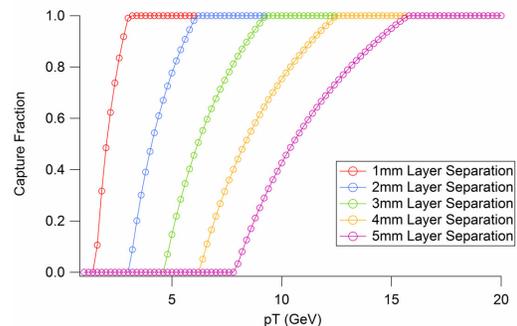

Figure 5: Capture probabilities for particles with varying transverse momenta. The rφ is 20µm. Inner sensor radius is 10cm.

The range over which the transverse momentum is cut depends on several factors. Increasing the layer separation and the radial position of the stack increases the $p_T$ at which the particles are cut, whereas increasing the size of the search window reduces it. The pixel size in $r\varphi$ determines the range over which the transverse momentum may or may not be cut (this is a by-product of the binary readout).

## II. DESIGN CONSIDERATIONS

In order to implement the system described above, there are three key design areas that need to be considered. These are the sensors themselves, the correlator logic implementation and the mechanical aspect of the design.

### A. Sensor Design

#### 1) Sensor Type

There are several new sensor types approaching maturity that offer comparable or better performance than the current hybrid pixel systems. At a radius of 10cm from the interaction point the required radiation tolerance is $10^{16}$p/cm$^2$, 300Mrad. While this is challenging for the CMOS electronics, this has significant implications for the sensing element, as full depletion becomes impossible using thick sensors. Furthermore a charge collection speed of <5ns will be required if SLHC operates in an 80MHz bunch crossing mode.

The two sensor types discussed here are both monolithic approaches. Monolithic Active Pixel Sensors (MAPS) [6] are now a relatively mature technology that relies on a thin (a few microns) epitaxial layer of p-type silicon, with CMOS electronics placed on top. The epi-layer acts as a potential well, and is responsible for providing signal charge to n-well diodes. This involves a relatively standard manufacturing process. However the lack of a uniform field in the epi layer results in slow charge collection and poor radiation tolerance. The second approach is the Thin Film on ASIC (TFA) [7] technology, which requires a non-standard process (PCVD) to deposit the sensor material on top of a standard CMOS ASIC. Its key advantage is the separation of the sensor technology from the readout electronics. Furthermore the sensing material can be changed if new ones become available. Possible sensor materials in this case include a-Si:H, HgI and CdZnTe.

#### 2) Pixel Size

The pixel size for a stacked pixel detector is driven by several requirements. Firstly the pitch needs to be small enough to ensure low occupancy; however this is actually not important in this design and is easily achievable in current pixel processes. The real driver for a stacked pixel is the required detector resolution and the chosen transverse momentum cut. The requirement for SLHC is derived from matching the resolution of a stub produced in the pixel stack to the trigger tower size in the CMS calorimeter [8]. This yields a maximum $\Delta\eta\times\Delta\varphi$ of 0.087x0.087. As the $p_T$ of a charged particle track cannot be inferred by a single stack alone (because of close proximity of the two pixel layers has a deleterious effect on the $p_T$ resolution), an assumption must be made about the $p_T$ of the track in order to achieve the required $\Delta\varphi$ resolution (see Figure 7). The requirement for $\Delta\eta$ is dominated by the pixel detector resolution, and can be tuned to match the calorimeter window. The method used to calculate the stub resolution, $\Delta\eta$, is shown in Figure 6. The results for a pixel size of 20µmx50µmx10µm are shown in Figures 7 & 8. The results yield an approximate resolution of 0.05x0.08 for a $p_T$ greater than 20 GeV and a layer separation of 2mm.

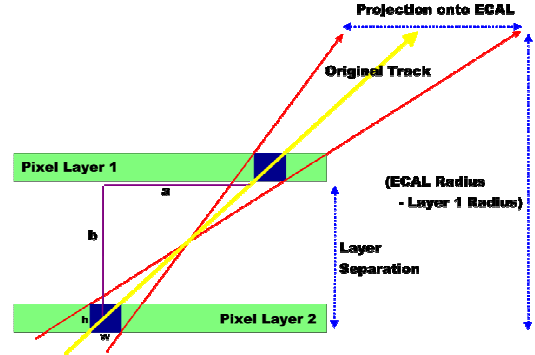

Figure 6: Minimum and maximum pseudorapidities for a given pixel pair. This is referred to as the min-max range. A similar method is used to calculate the $\Delta\varphi$ resolution.

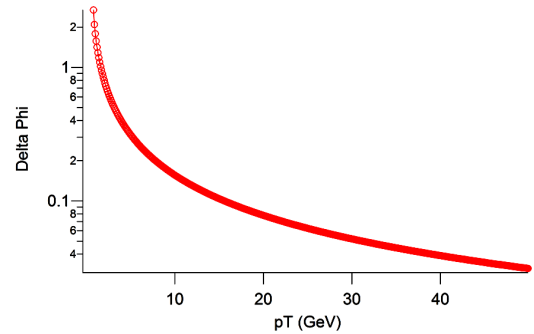

Figure 7: Angle in radians between the projected tangent of a track at its point of intersection with the stacked tracker and the point on the calorimeter which it hit, for a given particle $p_T$.

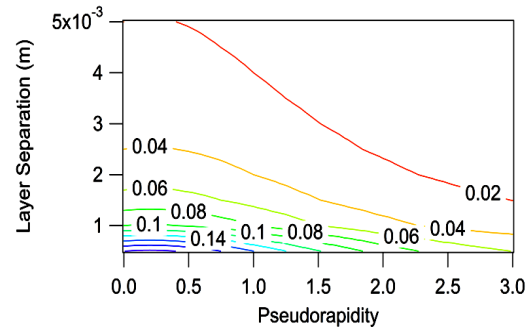

Figure 8: The stub resolution for a track extrapolated to the calorimeter. The values depend on both the separation between the two sensor layers and the position of the calorimeter hit. The values shown on the plot represent $\Delta\eta$.

*3) Pixel Readout Architecture*

In order for a pixel system to contribute to the L1 trigger, it is necessary for the digital bandwidth to be able to sustain the hit rates expected at SLHC. The new system will not be able to scan the pixels for hits quickly enough. Therefore one requires a self-triggering pixel design, which necessitates the use of an in-pixel comparator. The implementation of an analogue readout has been dismissed for this study.

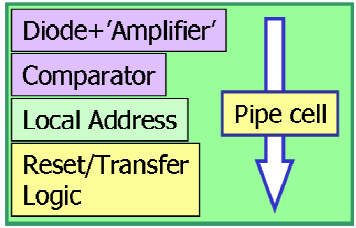

Figure 9: Basic design of a single pixel cell

The outline of a single pixel cell is shown in Figure 9. It is based on a pipelined column-parallel readout architecture where each pixel in a column forms a single cell in the pipeline, capable of storing a single hit address for that column. The architecture discussed here was born of two principles – the minimisation of power consumption and the maximisation of readout speed. In order to achieve this, a hit-scanning or token-based system with a global clock was dismissed in favour of a self-timed (asynchronous) system with self-triggering pixels. The reason for the dismissal of the global clock is two-fold. A global clock in this case would have to operate at speed of several GHz in order to transfer the data out of the sensor fast enough. Controlling the skew and stability of this system would involve very careful design and result in high power consumption. It would also consume a large amount of pixel real-estate for the implementation of clock buffers. The second reason for the dismissal of a synchronous clock design is because of logic simplicity. The readout requires a column-wise pipeline, where every pixel contains one of the pipeline cells in the chain. This can be implemented very easily using inverters in asynchronous design. In a synchronous mode, D-flip-flops would be required to avoid race conditions. There is simply not enough room in the pixels for these registers, and again they consume a large amount of power. There are several architectures that can be chosen for asynchronous micro-pipelines; one example of a four-phase pipeline is described in [9].

After the readout of the row address of the hit pixels is read out from a column, the column address is attached to the data. The hit addresses from all the columns are then concatenated into a column-ordered list (ordering of the columns is critical to the operation of the correlator, as will become clear in the next section). Due to the variable-size data blocks produced in this system, it is necessary to attach a timestamp to the front of the data-block to mark its position in the stream. The unfiltered data is then read out from the chip using a high-speed differential link at approximately 3.2Gb/cm$^2$/s. This high-speed link cannot be easily avoided as it is a consequence of the hit rate in the detector. The data is fed into the correlation ASIC along with that from the other sensor in the stack and the data is combined and filtered.

## B. Correlation Logic Implementation

As the data produced in the sensors is column-ordered with the lowest column address first, the implementation of the correlation logic is simplified to a simple difference analysis as follows. We have described the algorithm here in pseudo-C style code to simplify its presentation. The value $c_2$ denotes the column address of the currently considered piece of data in the outer sensor, whilst $c_1$ denotes that value for the inner sensor. The size of the search window is defined by x; for x=1 the search window would be nearest-neighbour.

**if ($c_2 > c_1 + x$) Next($c_1$)**

**else if ($c_1 > c_2 - x$) Next($c_2$)**

**else { Copy($c_1$, $c_2$); Next($c_1$) }**

The next bunch-crossing's-worth of data for the chip is marked by the timestamp at the front of the data block.

## C. Mechanical Design

The mechanical requirements of this design are a significant problem facing this approach. The reason for this is that any misalignment of the sensors with respect to the interaction point or to each other affects the performance of the system. Both of these complications can be calibrated against for off-detector processing.

*1) Inner Sensor-Outer Sensor Positioning*

The layout of the outer sensor with respect to the inner sensor in a stack has to be controlled precisely. The reason for this is that the highest $p_T$ tracks are assumed to traverse straight from the interaction point at r=0. One can align the central pixels in both sensors and simply accept that the lowest and highest column addresses in the outer sensor will not quite correspond to those in the inner sensor. This means that the $p_T$ cut will become both location-dependent and charge-dependent (but this may not be a serious issue). Alternatively for the pixel addresses to match in both the inner and outer sensor, the pixel pitch must be slightly larger in the outer layer (see Figure 10).

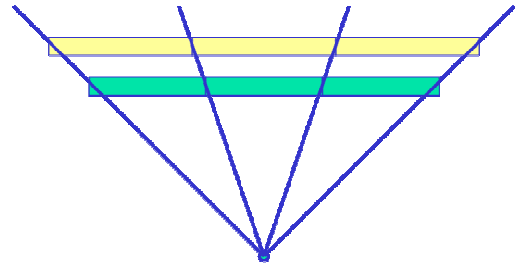

Figure 10: Illustration of different pitches in the inner and outer sensor. Note that those in the outer layer must be slightly larger than those in the inner layer.

*2) Stack-IP Positioning*

The relationship of the stack to the interaction point is also important. If the normal line of the stack is not parallel to the radial line coming from the interaction point, the stack will cut at a different $p_T$ for positively and negatively-charged particles. Again this may not be a serious problem for a system contributing to a Level-1 Trigger, but has not yet been quantified.

## III. MONTE-CARLO RESULTS

In order to gain an impression of the performance of this system, a simple Monte-Carlo simulation was developed to simulate the rate reduction in the detector. The data used was the same as that used by the full CMS simulation software. Minimum bias and Higgs → 4 lepton events were generated using Pythia 6.2772 via CMKIN 4.2. The simulation included a basic model of charge sharing and threshold triggering of the comparator, but did not include full energy deposition simulations and the more complex detector effects such as hadronisation and pair production. Nevertheless it is a useful simulation to illustrate the basic principles. A cross-section of the results are shown in Table 1, for a superposition of 200 minimum bias events (i.e. $10^{35}$@40MHz bunch crossing). The principle was tested for radial stack positions of r=10cm and r=20cm. The motivation for an r=20cm location is two-fold: firstly there is currently a space in the CMS tracker at this radius, where it may be possible to implement a new system without affecting the rest of the CMS tracker. Secondly one gains a rate and power density reduction of a factor of four simply because of the larger surface area of the detector.

Table 1: Readout data rate as a percentage of the unfiltered rate for 1-2mm layer separations at r=10cm and r=20cm. These figures depend significantly on the thresholds and charge sharing properties of the sensor, and so should only be considered approximate.

| Layer Separation | Radius (cm) | Readout Rate (%) |
|---|---|---|
| 1mm | 10 | 12.2 |
| 1mm | 20 | 3.19 |
| 2mm | 10 | 5.77 |
| 2mm | 20 | 1.68 |

The smallest rate reduction is naturally at the smallest radius and layer separation, as this represents the smallest $p_T$ cut out of those shown. As the number of charged particle tracks increases rapidly at low $p_T$, so does the corresponding rate reduction. For a radius of r=20cm, one also gains a factor of four in rate reduction per unit area simply because the detector is larger.

In a later test a high-$p_T$ lepton from the H→4l dataset was introduced into the event sample to verify it was detected every time. As expected the signal efficiency was 100%, which is a necessary requirement for this system to be useful in the Level-1 Trigger.

## IV. SUMMARY

It has been shown that the use of small layer separations in a pixellated detector system can be used to both reduce tracker combinatorials and reduce the data rate from the detector by means of a simple correlation algorithm. This algorithm could be implemented on-detector using relatively simple electronics; more advanced algorithms could be implemented off-detector in FPGAs.

The choice of sensor material is still undecided, and it is difficult to judge which material will be optimal for the final system. However the logic design can be investigated now.

Our future work involves further simulation using the full CMS detector simulation (OSCAR), with a modified geometry which includes a stack at r=20cm. We will also look at the mechanical requirements and their effect on track reconstruction in more detail. We then intend to investigate its use in the reconstruction algorithms at the Level-1 and Higher-Level Triggers, and the implementation of these algorithms in FPGAs.